\documentclass[12pt,twoside]{article}
\usepackage{fleqn,espcrc1}
\usepackage{graphicx}
   \newcommand{\cent}[1] {\begin{center}#1\end{center}}

   \newcommand{\lra}  {$\leftrightarrow$}
   \newcommand{\vecb}[1]{\mbox{\bf#1}}

\newcommand{\AmS}{{\protect\the\textfont2
  A\kern-.1667em\lower.5ex\hbox{M}\kern-.125emS}}

\title{Phase Transitions in ``Small'' Systems\\ -- A Challenge for 
Thermodynamics --\footnote{Invited talk at CRIS 2000, 3rd Catania
Relativistic Ion Studies ``Phase Transitions in Strong Interactions: 
Status and Perspectives'', Acicastello, Italy, May 22-26, 2000}}
\author{D.H.E. Gross\address{Hahn-Meitner-Institut
  Berlin, Bereich Theoretische Physik,Glienickerstr.100\\ 14109
  Berlin, Germany and Freie Universit{\"a}t Berlin, Fachbereich
  Physik}}
\begin{document}

\maketitle
\begin{abstract} Traditionally, phase transitions are defined in the
  thermodynamic limit only. We propose a new formulation of equilibrium
  thermo-dynamics that is based entirely on mechanics and reflects
  just the {\em geometry and topology} of the N-body phase-space as
  function of the conserved quantities, energy, particle number and
  others. This allows to define thermo-statistics {\em without the use
    of the thermodynamic limit}, to apply it to ``Small'' systems as
  well and to define phase transitions unambiguously also there.
  ``Small'' systems are systems where the linear dimension is of the
  characteristic range of the interaction between the particles. Also
  astrophysical systems are ``Small'' in this sense.  Boltzmann
  defines the entropy as the logarithm of the area $W(E,N)=e^{S(E,N)}$
  of the surface in the mechanical N-body phase space at total energy
  $E$.  The topology of $S(E,N)$ or more precisely, of the curvature
  determinant $D(E,N)=\partial^2S/\partial E^2*\partial^2S/\partial
  N^2-(\partial^2S/\partial E\partial N)^2$ allows the classification
  of phase transitions {\em without taking the thermodynamic limit}.
  The topology gives further a simple and transparent definition of
  the {\em order parameter.}  Attention: Boltzmann's entropy $S(E)$ as
  defined here is different from the information entropy c.f.
  \cite{balian91} and can even be non-extensive and convex.
\end{abstract}
\section{Fundamentals of thermo-statistics \label{fundamental}}
Conventional (canonical) thermo-statistics addresses large (in the
thermodynamical limit), homogeneous systems. Extensivity (i.e. if the
system is divided into pieces their energy and entropy scale with the
size of the pieces) is an important condition c.f.\cite{lieb97}.

Here we propose a new, easy, and more transparent access to the
thermodynamics and especially to phase transitions which applies also
to ``Small'' systems. Only the static geometrical and topological
properties of the volume of energy-shell of the N-body phase-space is
investigated. {\em No thermodynamic limit has to be invoked}. It is
obvious that only by this extension of thermo-statistics it is
possible to discuss phase transitions in nuclei, atomic clusters and
astro-physical systems.

First order transitions are distinguished from continuous transitions
by the appearance of phase-separations.  Here the system becomes
inhomogeneous and coexistent phases are separated by interfaces. Any
system at phase separation is necessarily {\em inhomogeneous and
  non-extensive}.  This is also the case for the majority of systems
in nature: hot nuclei, hot atomic clusters and the real big ones:
astrophysical systems under self-gravity.  They are inhomogeneous even
away from phase transitions. There the thermodynamic limit makes no
sense.  We will henceforth call these systems ``Small'' or
``non-extensive''.

To describe these non-extensive systems, we have to go back to
pre-Gibbsian times.  Boltzmann's famous epitaph
\cent{\fbox{\fbox{$S=k*lnW$}}} contains everything what can be said
about equilibrium thermodynamics in its most condensed form. $W$ is
the volume of the sub-manifold of sharp energy in the $6N$-dim.  phase
space.  It defines entropy $S$ and with it thermodynamics entirely by
mechanical quantities and geometry.  No thermodynamic limit, no
extensivity, no concavity (downwards bending) of the entropy have to
be invoked. This was largely forgotten since hundred years.

\section{Thermo-statistics as geometry and topology of the mechanical
  phase-space} Like Boltzmann we are led by mechanics as the safe
guide when we want to extend thermodynamics here to non-extensive or
``Small'' systems.  For ``Small'' systems the canonical ensemble and
with it the Boltzmann-Gibbs distribution have no support by mechanics
anymore.  A fact by the way Gibbs agreed fully with \cite{gibbs02f}.
Whereas Boltzmann's definition above opens the way to interprete
thermodynamics by the {\em topological and geometrical} properties of
the N-body phase space. This is new and in marked contrast to e.g.
the book of Balian \cite{balian91} where the entropy is derived from
information theory and its extensivity as well its concavity seems to
be demanded from the beginning. Boltzmann's definition as written
above is free of this. It is a {\em purely geometric} definition and
therefore simple. As it allows for convexity as well for
non-extensivity without contradicting the Second Law of thermodynamics
\cite{gross173} it is much more suited for the purpose of
non-extensive thermo-statistics of ``Small'' systems.

Another fairy tale told to us by most textbooks of statistical
mechanics and thermodynamics says: ``Phase transitions exist only in
the thermodynamic limit.'' There, we also learn that phase transitions
get smeared in finite systems. Nevertheless though, phase transitions
do exist in these systems. It was the result of experiments and
theoretical thoughts especially in {\em nuclear physics} that paved
the way to a new and much deeper understanding of equilibrium
statistical mechanics and phase transitions in ``Small'' systems. I
want to remind the pioneering attempts on nuclear multifragmentation,
experimentally by the references
\cite{kaufmann80,jakobsson82,friedlander83,moretto84,aichelin86,klotz87,ogilvie91,hubele91,blumenfeld91},
where finally Moretto \cite{moretto92a} gave a concluding overview on
the experimental situation of multifragmentation, even though he
initially doubted the very existence of this new phenomenon in the
paper entitled: ``Complex fragment emission at 50 MeV/u - compound
nuclei for ever'' \cite{bowman87}. This development was soon
theoretically accompanied and stimulated:
\cite{gross45,bondorf81,bondorf82,gross58,gross69,botvina85,campi88,peilert88,lopez89,chung89,gross95,bondorf95,gross153}
and others. In fact the paper \cite{gross58} is the first of some 500
papers on ``multifragmentation'' cited in the large citation database
ISI ``Web of Science''. Very early, multifragmentation was discussed
as a real phase transition of first order in nuclei. E.g. in
\cite{gross69} it was argued that multifragmentation is a transition
of first order distinct from the usual liquid-gas transition. And of
course this was violently attacked also. I remember some experimental
colleagues who claimed the intermediate mass fragments were coming
from the silicon grease used in the vacuum chamber. Theoretically, it
was warned that phase transitions cannot occur in such small systems
like nuclei. They will be washed out and maybe not even recognizable.
In contrast to phase transitions discussed so far in nuclear physics
like the transition from spherical to deformed nuclei or the one from
normal to superfluidity which are all entropy $S=0$ phenomena, here
the real macroscopic signals of a transition in phase-space are seen
where the {\em entropy shows some anomaly}: A sudden change of the
configuration, a latent heat, even a negative heat capacity is seen
c.f.  \cite{gross69,gross95,bondorf95,pochodzalla95,gross153} which
is linked to a convexity of the entropy, forbidden in normal
extensive thermodynamics by van Hove's concavity rule
\cite{vanhove49}.
\section{The whole ``zoo'' of phase transitions exists in ``Small'' systems
\label{zoo}}
\begin{figure}[h]
  \includegraphics*[bb =0 0 290 180, angle=0, width=15cm,
clip=true]{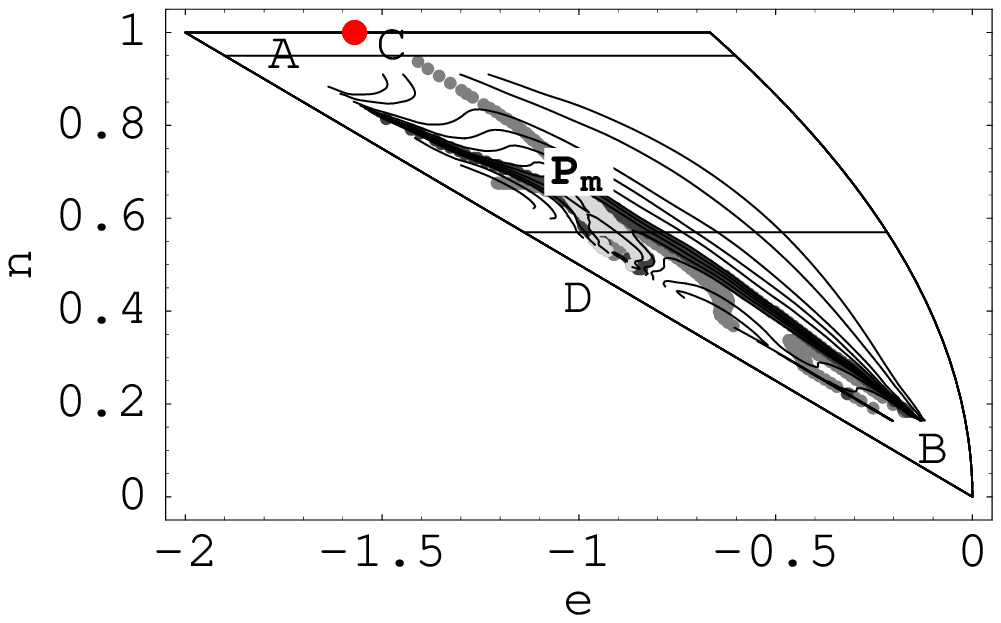}
\\
\caption{Contour plot of the determinant of curvatures
  $\det(e,n)$ defined in eq.(\ref{curvdet}) of a ($q=3$) Potts lattice
  gas on a $50*50$ lattice. Regions above $\widehat{CP_mB}$ : concave,
  $\det>0,\lambda_1<0$ (we always have $\lambda_2<0$), pure phase
  (disordered, gas), in the triangle $A$$P_m$$C$ concave, pure phase
  (ordered, solid); below $\widehat{AP_mB}$: convex,
  $\det<0,\lambda_1>0$, phase-separation, first order; At the dark
  lines $\widehat{AP_mB}$ we have $\det(e,n)=0,\lambda_1=0$:
  termination lines of the first order transition; Medium dark lines
  e.g $\widehat{CP_m}$.:
  $\vecb{v}_1\cdot\mbox{\boldmath$\nabla$}\det=0$; here the curvature
  determinant has a minimum in the direction of the largest curvature
  eigenvector $\vecb{v}_1$; in the cross-region (light gray) we have:
  $\det=0\wedge\mbox{\boldmath$\nabla$}\det=\mbox{\boldmath$0$}$ this
  is the locus of the multi-critical point $P_m$. In a {\em two}
  dimensional neighborhood of which $s(e,n)$ is flat in in the
  direction of the largest curvature and downwards curved in the other
  (like the surface of a cylinder) up to at least third order of
  $\Delta e$ and $\Delta n$ and $\det(e,n)=0$. The two horizontal
  lines give the positions of the two cuts shown in fig.(\ref{cuts})
\label{det}}
\end{figure}

\begin{figure}[h]
\begin{minipage}[t]{7cm}
  \includegraphics*[bb =0 0 290 180, angle=-0, width=7cm,
  clip=true]{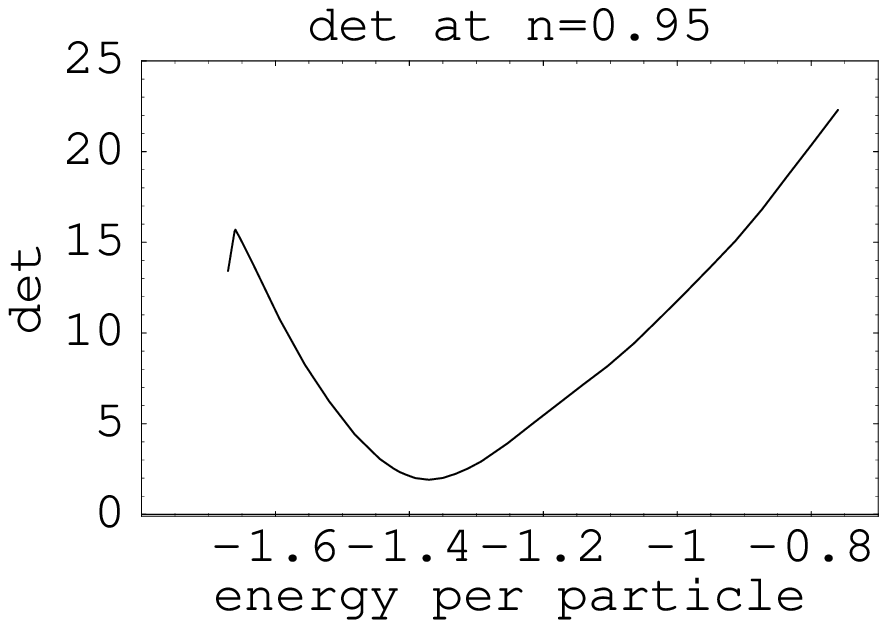}
  \\
  Cut through the determinant $\det(e,n)$ along the line shown in
  figure (\ref{det}) at const.  $n=0.95$, through the critical line
  $\widehat{CP_m}$ close to the critical point $C$ of the ordinary
  Potts model ($n\sim 1$)
\end{minipage}\rule{0.5cm}{0mm}\begin{minipage}[t]{7cm}
  \includegraphics*[bb =0 0 290 180, angle=-0, width=7cm,
  clip=true]{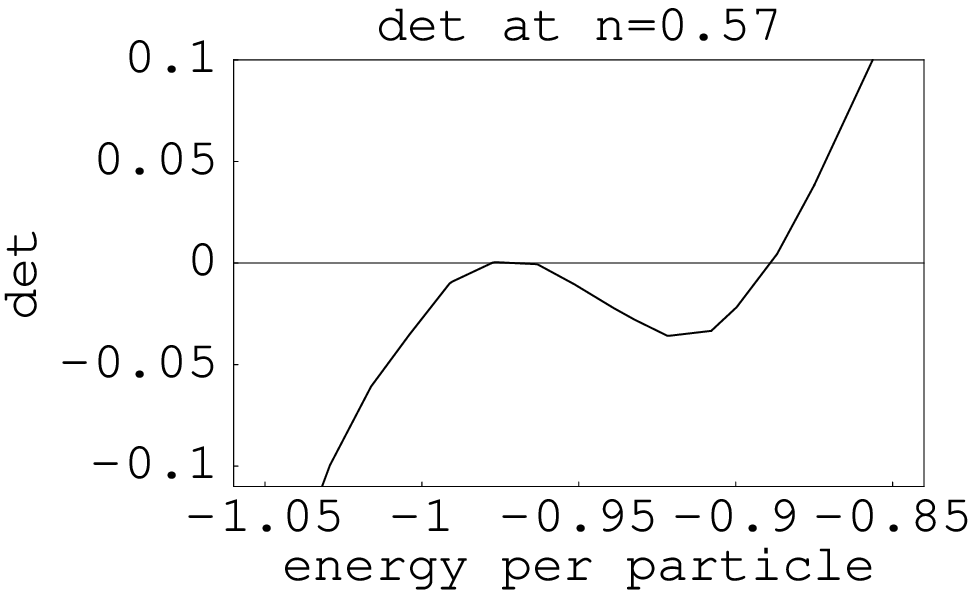}
  \\
  Cut through the determinant $\det(e,n)$ along the line shown in
  figure (\ref{det}) at const.  $n=0.57$, slightly below the
  multi-critical region. There are several zero points of the
  determinant of curvatures: The left one is simultaneously a maximum
  with $\mbox{\boldmath$\nabla \det=0$}$ and consequently critical as
  discussed above
\end{minipage}\caption{\label{cuts}}\end{figure}

\begin{figure}
    \includegraphics*[bb =0 0 290 180, angle=-0,
    width=11cm,clip=true]{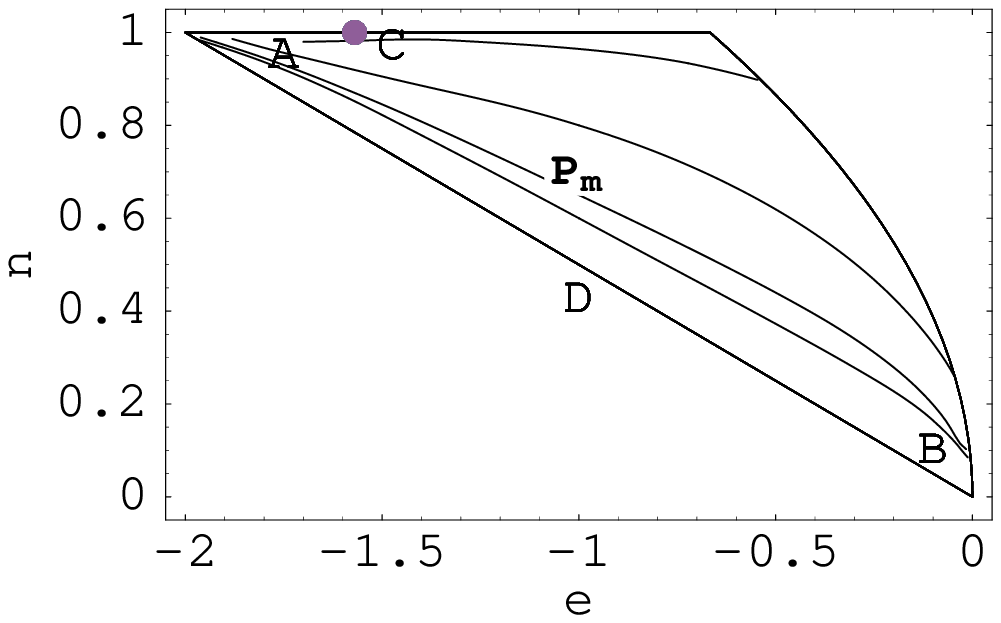}
\caption{Direction $\mbox{\boldmath$v_1$}$ of the largest principal curvature 
  $\lambda_1$ which defines the order parameter that tunes the system
  through the phase-transitions along these lines.
\label{maincurvature}}
\end{figure}

In this talk I will show just for demonstration how the whole ``zoo''
of phase transitions: first order transitions {\em including the
  interphase surface tension}, continuous transitions, critical and
even multi-critical points are unambiguously and {\em sharply} defined
in ``Small'' systems of some hundred particles {\em by the curvatures
  of the micro-canonical entropy surface $s(e,n)$.}  Details can be
found in \cite{gross173,gross174}. For an example, I show
(figs.\ref{det} and \ref{cuts}) here only the phase diagram of the
determinant of curvatures of $s(e,n)=S/V$ vs.  energy per volume
(lattice point) $e=E/V$ and density (occupation) $n=N/V$ for a Potts
lattice gas of $50*50$ lattice points with three spin components
($q=3$):
\begin{equation}
\det(e,n)= \left\|\begin{array}{cc}
\frac{\partial^2 s}{\partial e^2}& \frac{\partial^2 s}{\partial n\partial e}\\
\frac{\partial^2 s}{\partial e\partial n}& \frac{\partial^2 s}{\partial n^2}
\end{array}\right\|
= \left\|\begin{array}{cc}
s_{ee}&s_{en}\\
s_{ne}&s_{nn}
\end{array}\right\|=\lambda_1\lambda_2,\hspace{1cm}\lambda_1\ge\lambda_2
 \label{curvdet}
\end{equation}
~\\
The eigenvectors of the curvature matrix define the main curvature
directions. Notice, the smaller of the two eigenvalues is here always
$\lambda_2<0$. The larger eigenvalue $\lambda_1$ may be positive or
negative. The direction of the eigenvector $\vecb{v}_1$ belonging to
$\lambda_1$ gives a natural and unambiguous definition of the {\em
  order parameter} of the system.  It is by a progression in this
direction that the system changes from one phase over a region of
phase-separation to the other phase. In figure (\ref{maincurvature})
we see the phase transition of first order is controlled by variation
roughly parallel to the ground-state (similar to the magnetization
axis in the Ising model) and the phase transition of second order in
the pure Potts-model by a variation $\sim$ the energy. As can be
well seen the order parameter is not always a straight line in the
parameter space.
The systematics of micro-canonical phase transitions will be listed in
the conclusion.  The multi-critical point $P_m$ is most interesting:
Here the largest curvature eigenvalue $\lambda_1$ and therefore also
the curvature determinant $\det(e,n)$ vanishes in a {\em
  two-dimensional} neighborhood.  The entropy surface $s(e,n)$ has the
topology of the surface of a cylinder inside this neighborhood.

As is discussed in all details in ref.\cite{gross173,gross174} this
new and more extended definition of phase transitions agrees with the
Yang-Lee definition in the thermodynamic limit but it allows to define
phase transition also in the much larger world of non-extensive
systems.  {\em Applications of thermodynamic arguments to these
  systems and of course to hot nuclei make sense only within the
  micro-canonical theory.} This is also a serious reminder for many
contributions to this conference.  Moreover, thermodynamics of
``Small'' systems like hot nuclei is naturally probabilistic and
therefore requires normally a Monte Carlo kind of treatment.
\section{Conclusion \label{concl}}
Micro-canonical thermo-statistics describes how the entropy $s(e,n)$
as defined entirely in mechanical terms by Boltzmann depends on the
conserved ``extensive'' variables: energy $e$, particle number $n$,
angular momentum $L$ etc.  In contrast to the conventional theory, we
can study phase transitions also in ``Small'' systems or other
non-extensive systems. They are sharply defined for finite systems
without invoking the thermodynamic limit.  We classify phase
transitions in a ``Small'' system by the topological properties of the
determinant of curvatures $\det(e,n)$, eq.(\ref{curvdet}), of the
micro-canonical entropy-surface $s(e,n)$:
\begin{itemize}
\item A single stable phase by $\det(e,n)>0$. Here $s(e,n)$ is
concave (downwards bended) in both directions and there is a one to
one mapping of \{e,n\} \lra \{T,$\nu$\} or between the micro and the
grand-canonical ensemble.
\item A transition of first order with phase separation by
  $\det(e,n)<0$ or the largest curvature $\lambda_1>0$.  Here the
  entropy surface $s(e,n)$ has a convex intruder.  The depth of the
  intruder is a measure of the inter-phase surface tension
  \cite{binder82,gross150,gross157}. This region is bounded by a line
  with $\det(e,n)=0$.
\item On this line $P_m$ is a critical end-point where additionally
  $\vecb{v}_1\cdot\mbox{\boldmath$\nabla$} \det=0$ in the direction of
  the eigenvector of $\det(e,n)$ with the largest eigenvalue
  $\lambda_1$. I.e. $\det(e,n)$ has here a minimum.  There, the
  transition is continuous (``second order'') with vanishing surface
  tension, and no convex intruder in $s(e,n)$. Here two neighboring
  phases become indistinguishable, because there are no interfaces.
  Moreover, we found a further {\em line} ($\widehat{P_mC}$, critical)
  with $\vecb{v}_1\cdot\mbox{\boldmath$\nabla$} \det=0$ which does not
  border a region of negative $\det(e,n)$.  Presumably $\det(e,n)$
  should be $0$ also.  This needs further tests in other systems. It
  may further be that these lines signalize transitions of first order in
  another, but hidden non-conserved order parameter, e.g. the staggered
  magnetization c.f.\cite{gross174}.
\item Finally a multi-critical point $P_m$ where more than two phases
  become indistinguishable by the branching of several lines with
  $\det=0$ or with $\vecb{v}_1\cdot\mbox{\boldmath$\nabla$} \det=0$ to
  give a {\em cylindrical} region of $s(e,n)$ with additionally
  $\mbox{\boldmath$\nabla$} \det=\mbox{\boldmath$0$}$, here the
  largest curvature $\lambda_1 =0$ has a maximum in a 2-dim.
  neighborhood.
\item All regions with $\det(e,n)\le 0$ lead to the catastrophes of
  the Laplace transform from the micro to the grand-canonical ensemble
  and thus to the Yang-Lee singularities of the grand-canonical
  partition sum in the thermodynamic limit.
\end{itemize}

It is fair to say that the discovery of nuclear multi-fragmentation as
a real phase transition of first order in a {\em``Small''} many-body
system is a challenge for statistical mechanics to understand its
foundation better and to become able to describe also the
thermodynamics of inhomogeneous and non-extensive systems
\cite{gross175}. It opens thermo-statistics for so many applications
from small systems like nuclei up to the largest like astro-physical
ones.  Boltzmann's definition of entropy allows for a {\em geometrical
  and topological} interpretation of thermo-statistics with the virtue
of great conceptional clarity and to be free of invoking the thermodynamic
limit. Nuclear collisions are a beautiful testing ground of these new
ideas.  This is of course an idealized description.  Dynamical
non-statistical features are certainly also there.

\end{document}